\documentclass{emulateapj}

\usepackage{natbib}
\usepackage{amsmath}
\usepackage{gensymb}
\shorttitle{Palomar 5}
\shortauthors{FRITZ \& KALLIVAYALIL} 

\begin{document}
 \title{The proper motion of Palomar 5\altaffilmark{$*$}}

\author{ T.K.~Fritz\altaffilmark{1,$\#$}, N.~Kallivayalil\altaffilmark{1}
 }

 \altaffiltext{$*$}{Based on LBT data. The LBT is an international collaboration among institutions in the United States, Italy and Germany. LBT Corporation partners are: 
 The Ohio State University, and The Research Corporation, on behalf of The University of Notre Dame, University of Minnesota and University of Virginia; LBT Beteiligungsgesellschaft, Germany, representing the Max-Planck Society, the Astrophysical Institute Potsdam, and Heidelberg University; the University of Arizona on behalf of the Arizona university system; and Istituto Nazionale di Astrofisica, Italy.
 }
\altaffiltext{1}{Department of Astronomy, University of Virginia, Charlottesville, 3530 McCormick Road, VA 22904-4325, USA}
\altaffiltext{$\#$}{E-mail: tkf4w@astro.virginia.edu}


\begin{abstract}
Palomar 5 (Pal~5) is a faint halo globular cluster associated with narrow tidal tails. It is  a useful system to understand the process of tidal dissolution, as well as to constrain the potential of the Milky Way. A well-determined orbit for Pal~5 would enable detailed study of these open questions.
We present here the first CCD-based proper motion measurement of Pal~5 obtained using SDSS as a first epoch and
new LBT/LBC images as a second, giving a baseline of 15 years.
We perform relative astrometry, using SDSS as a distortion-free reference, and images of the cluster and also of the Pal~5 stream for the derivation of the distortion correction for LBC.
The reference frame is made up of background galaxies.  
We correct for differential chromatic refraction using relations obtained from SDSS colors as well as from flux-calibrated spectra, finding that the correction relations for stars and for galaxies are different.  
We obtain $\mu_\alpha=-2.296\pm0.186$ mas/yr and $\mu_\delta=-2.257\pm0.181$ mas/yr for the proper motion of Pal~5.
We use this motion, and the publicly available code \textit{galpy}, to model the disruption of Pal~5 in 
 different Milky Way models consisting of a bulge, a disk and a spherical dark matter halo. Our fits to the observed stream properties (streak and radial velocity gradient)  result in a preference for a relatively large Pal~5 distance of around 24 kpc. A slightly larger absolute proper motion than what we measure also results in better matches but the best solutions need a change in distance. 
 We find that a spherical Milky Way model, with $V_0=220$ km/s and  V$_\mathrm{20\,kpc}$, i.e., approximately at the apocenter of Pal~5, of 218 km/s, can match the data well, at least for our choice of disk and bulge parametrization.
\end{abstract}
\maketitle

\section{Introduction}
\label{sec:intro}

The globular clusters of the Milky Way have a range of luminosities \citep{Harris_96}, and Palomar 5 (Pal~5) \citep{Abell_55} is one of the faintest. It is located in the halo, at a distance of $\sim22$ 
kpc from the sun, and is associated with thin tidal tails of $\geq22\degree$ in length \citep{Odenkirchen_01,Grillmair_06}.
 Pal~5 shows unambiguously that tidal disruption is an important process for globular clusters,
 and offers the possibility to study tidal disruption in detail at a crucial phase. 
Other globular clusters
such as GD-1 appear to have already disrupted to thin tails without a clear parent object \citep{Grillmair_06b,Grillmair_14}. At least 17\% of halo stars show chemical signs  of once having been members of globular clusters \citep{Martell_11}. Thus globular clusters are an important contributer to the stellar halo. According to \citet{Dehnen_04}, Pal~5 will likely be totally disrupted during its next disk passage. However, the orbit of Pal~5 is currently not known well enough to strongly constrain its disruption time.

Streams are useful in determining the shape of the dark matter halo because the stars in the tails, once  stripped from the progenitor, feel negligible influence from this parent body, and can be used as test particles in determining the underlying Milky Way potential  \citep[e.g.][]{Johnston_99,Majewski_03,Law_10,Koposov_10,Bonaca_14,Price_14}. 
Therefore,  Pal~5 and its stream offer the possibility to constrain the potential of the Milky Way \citep{ Odenkirchen_03,Odenkirchen_09,Pearson_14,Kuepper_15}.
However, existing work on various Milky Way streams offer no real conclusion about halo shape. For instance, even the well-studied Sagittarius stream has been used to argue for different halo shapes:
spherical \citep{Ibata_01}, oblate \citep{Johnston_05}, prolate \citep{Helmi_04}, and triaxial \citep{Law_09}.
\citet{Law_10} fit data
from the Sagittarius stream, obtaining a slightly triaxial Milky Way halo
(not outside the expectations of $\Lambda+$cold dark matter-based structure formation; hereafter $\Lambda$CDM), but with 
a minor axis that lies within the Galactic disk.
Such a misalignment is difficult to reconcile with galaxy formation models and is expected to be unstable due to torques (\citealt{Debattista_13}; see also \citealt{Vera_13}).
 Modeling of other streams usually results in a preference for a spherical 
 or slightly oblate halo \citep{Koposov_10,Bowden_15,Kuepper_15}.
Further tests on the shape of the halo are thus warranted. 

Streams are also useful tools for measuring the enclosed mass of the Milky Way \citep[e.g.][]{Gibbons_14}.
The Milky Way mass is still poorly constrained. 
\citet{Gibbons_14}
obtain a value of $6\times10^{11}$ M$_\odot$ using the Sgr stream, nearly a factor of three less than the mass obtained by \citet{Boylan_13}: $16\times10^{11}$ M$_\odot$ using Leo I;  see \citet{Wang_15} for a recent compilation of Milky Way mass measurements.
Due to the fact that it is relatively close-by, Pal~5 can also be used to constrain the Milky Way disk. 

 Finally, Pal~5 shows tantalizing gaps in its stream \citep{Carlberg_12b}. These may be caused by substructure \citep{Dehnen_04}, like  giant molecular clouds, spiral arms or dark subhalos \citep{Carlberg_12a}, or by the epicyclic nature of tidal disruption even in a smooth halo \citep{Kuepper_08,Just_09,Kuepper_12,Mastrobuono_12,Kuepper_15}.
Gaps are well detectable only in thin (cold) streams, which makes Pal~5 useful to study substructure \citep{Yoon_11,Carlberg_12b}.

To shed light on all of these topics a proper motion measurement would be useful. A proper motion in combination with the known radial velocity of the cluster gives Pal~5's present-day kinetic energy directly, which enables better orbit constraints, and in turn, fewer free parameters in modeling the tidal disruption of the system.
Currently, there are three photographic plate-based proper motion measurements  \citep{Dinescu_99} which contradict each other. 
In this work, we obtain the first CCD-based proper motion of the Pal~5 cluster, using two large field-of-view imagers, SDSS and the Large Binocular Camera.
In Section~\ref{sec:dataset} we present the data. We describe the measurement of the proper motion of Pal~5 in Section~\ref{sec:deriv_pm}.  In Section~\ref{sec:MW_mass} we use the obtained proper motion to explore the Milky Way potential and the orbit and distance of Pal~5.
We discuss and conclude in Section~\ref{sec:discussion}.

\section{Data Set}
\label{sec:dataset}

In this section we describe the data used to measure the proper motion of Pal~5.

\subsection{SDSS}
\label{sec:sdss}

We use SDSS data release DR9 \citep{Ahn_12} as our first epoch of data. Subsequent data releases do not contain new imaging. 
We retrieve all {\tt clean} objects in the Pal~5 stream area from the {\tt Photo} table. This  includes generous margins around the stream debris location, and also covers our second epoch imaging area.
Selecting {\tt clean} objects means that we exclude duplicates, sources with obvious deblending and/or interpolation
problems, spurious detections, and stars close to the edge.
The {\tt clean} criterion does not work perfectly; in addition we later exclude sources using other criteria, as discussed in Section~\ref{sec:selection}.
In the case of stars we use {\tt PSF} magnitudes and in the case of galaxies we use {\tt cmodel} magnitudes. 
SDSS uses {\tt cmodel} magnitudes to separate stars from galaxies. Stars ($\mathrm{ProbPSF}_\mathrm{band}=$1) are defined as 
{\tt PSF\,mag$_\mathrm{\,band}$-cmodel\,mag$_\mathrm{\,band}$}$
<0.145$. 
For the DCR correction we use the measured magnitudes, see Section~\ref{sec:dcr}.
For the purpose of selecting of Pal~5 stars (Section~\ref{sec:selection}), we use extinction-corrected magnitudes, relying on the SDSS-provided correction based on \citet{Schlegel_98}. We use the given SDSS positions and their errors as a base. We then iterate on these positions to better fit the differential chromatic refraction (DCR) correction as well as the error model, as discussed in Sections~\ref{sec:dcr} and \ref{sec:distortion}.
The epoch of the SDSS data is 1999, with an average \textit{MJD}$=51269$.

\subsection{LBC/LBT}
\label{sec:lbt}

\begin{figure}
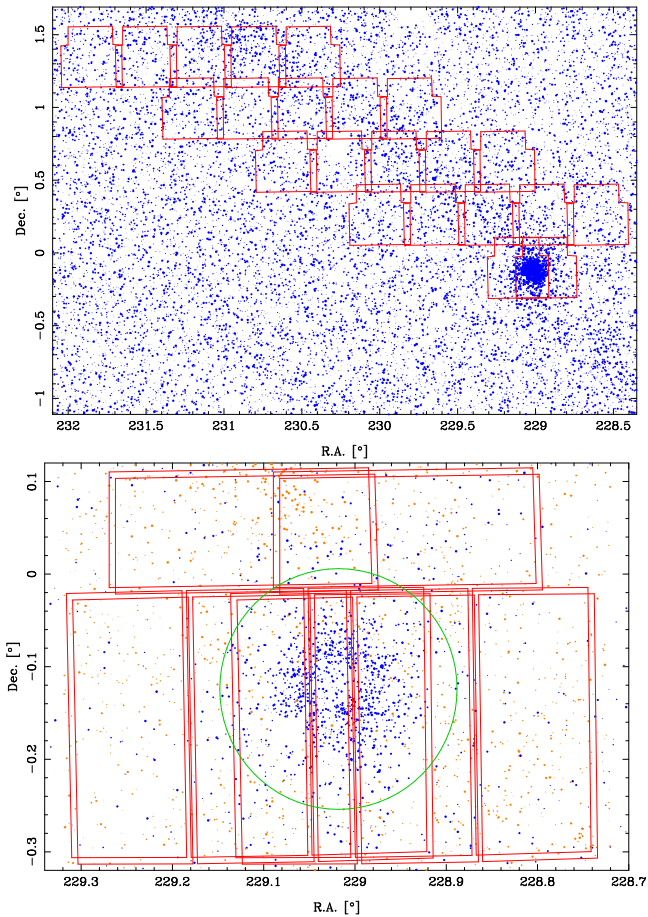

\begin{center}
\includegraphics[width=0.70 \columnwidth,angle=-90]{f1a.eps} 
\includegraphics[width=0.70 \columnwidth,angle=-90]{f1b.eps} 
\caption{ 
Map of LBC coverage of Pal~5. Top, the stream area: the blue dots show the stars which fall in the Pal~5 matched filter, and thus are likely members of the Pal~5 system. The area of the dots is proportional to the weight within the filter.  The cluster Pal~5 is around 229.02/-0.12.
The red polygons show the field of view of one LBC pointing. We cover the area with 22 different pointings. 
Bottom, the cluster itself: blue dots show stars in the matched filter, the central hole is caused by source blending in SDSS.
The green circle (r$=0.13\degree$) marks the outer boundary of the cluster as used in this work.
The red rectangles show the four detectors of the four different pointings on the cluster.
The orange dots show the reference galaxies. Their area is antiproportional to their position uncertainty.
} 
\label{fig:datdis}
\end{center}
\end{figure}

For the second epoch of imaging we use the Large Binocular Camera (LBC) \citep{Giallongo_08} at the Large Binocular Telescope (LBT).  
These data were obtained on 2014 July 1, giving a 15-year baseline between the first and second epoch of imaging. All LBC images have exposure times of 40 seconds and cover a field of view of about $23'\times25'$. The field is divided into four detectors, three upright at the bottom and one prone 
at the top. Each individual detector covers $7.8'\times 17.6'$. We obtained 12 images of the cluster and 41 images of the stream.
The images of the cluster overlap generously, while the images of the stream (usually two images per position) have only small overlaps between them, see Figure~\ref{fig:datdis}.
We use the $r$-band data of the red eye of the LBT. The r-band is the reddest of the high SNR LBC bands and is thus less influenced
by atmospheric effects such as DCR. 
The pixel scale of LBC is 0.226$\arcsec$. The typically achieved FWHM is $0.9\arcsec$.

We use standard data reduction techniques on the LBC data: we obtained skyflats, biases, and from the combination of the two, bad-pixel masks. We do not correct for cosmic rays or other artifacts (such as those around very bright stars) in the individual  images. Instead we use outlier rejection to exclude bad pixels in our source lists. Since we have between 2 and 12 images of each pointing, this is robust and does not reduce the number of sources in a relevant way. Stars close to very bright stars are usually already excluded from the {\tt clean} SDSS source list. Thus, for our current purposes there is little point in applying more advanced data reduction techniques to be able to extract more sources close to very bright stars from the LBC data.
We do not combine the individual LBC images into stacks, as position errors can be better estimated by using each of the individual images. Furthermore, image-stacking can often degrade the astrometry, when the images cover different pointings 
(especially when the pointing difference is only slightly smaller than the field of view of the instrument, see e.g. \citealt{Gillessen_09}), and when the pixel scale varies significantly over the field of view \citep{Giallongo_08,Bellini_10}.

\section{Deriving the proper motion of the Pal~5 cluster}
\label{sec:deriv_pm}

It is in principle possible to measure the proper motion of the stream from our data. However, due to the small contrast of the stream compared to the background (see e.g. Figure~\ref{fig:datdis} top) more aerial coverage is necessary than has been obtained here, and this will instead be the focus of subsequent work.
In this paper we focus on the measurement of the cluster proper motion. 

\subsection{Obtaining pixel positions}
\label{sec:pix_pos}

Since we use SDSS as our first epoch, we are constrained to use only SDSS-detected sources in our analysis. Starting with these as an input list, we use a preliminary version of our coordinate transformations (Section~\ref{sec:distortion})
to obtain the expected pixel positions of SDSS sources in our LBC images. We then search within a radius of 4 pixels ($\approx 0.9"$) for the local maximum\footnote{We use here for many basic steps: dpuser, see http://www.mpe.mpg.de/~ott/dpuser/index.html}. This search box is sufficient to find all objects except for high proper motion stars, which are foreground stars and not the targets of the present analysis.
We then fit a two-dimensional Gaussian to the 4-pixel half-width box around that maximum, leaving both widths and the orientation of the major axis of the Gaussian as  free parameters. We exclude very small and large widths, which likely result from failed fits. At this point we do not discriminate between stars and galaxies. We use two variants of the LBC images, the standard reduced images, and images which we obtain by smoothing these with a Gaussian of FWHM$=4$ pixels. 
There are no systematic differences in the positions obtained from these two kinds of images. 
However, we use the positions obtained from the smoothed images, since in this case the fits fail more rarely. The fits fail only for $\sim6$\% of the sources, which are usually faint. The smoothing also reduces the difference in shape between our simple source models and the actual sources that in reality are more complex.
It is unlikely that our source shape model is limiting our accuracy. Our error floor (Section~\ref{sec:selection}) is consistent with the error floor of SDSS \citep{Pier_03}.
Also the total error is dominated by the SDSS errors, see Section~\ref{sec:vel}, which shows that our simple shape model is sufficient to obtain positions.

\subsection{Differential chromatic refraction correction}
\label{sec:dcr}

\begin{figure}
\begin{center}
\includegraphics[width=0.70 \columnwidth,angle=-90]{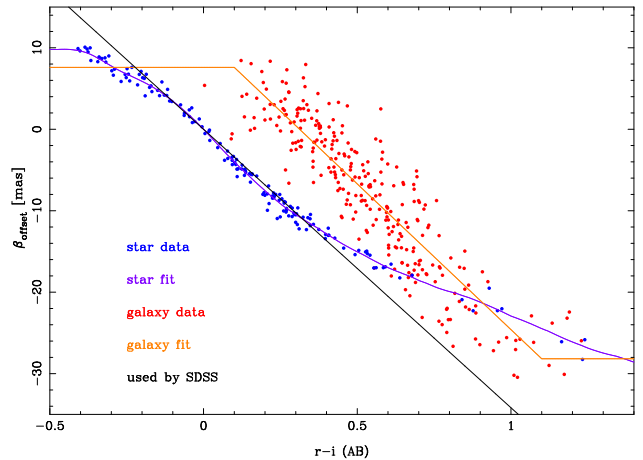} 
\caption{Differential chromatic refraction as function of $r-i$ color. $\beta_\mathrm{offset}$ is the shift to zenith for $\zeta=45\degree$. The shift for stars with $r-i=0$ is set to zero.
The black line is the relation used by SDSS. We obtain for stars a similar relation (violet line). However, our relation for galaxies (orange line) is clearly offset from this relation. 
} 
\label{fig:dcr1}
\end{center}
\end{figure}

The atmosphere of the earth deflects light with an angle of \begin{equation} \alpha=\alpha' \tan(\zeta) 
\end{equation} in which $\zeta$  is the angle from zenith and $\alpha'$ follows from the index of refraction, $n(\lambda)$:
\begin{equation}
\alpha'=\frac{n(\lambda)^2-1}{2\,n(\lambda)^2}.
\end{equation}
$\alpha$ is 57'' for $\zeta=45\degree$.
Most of the deflection, $\alpha$, is corrected for  automatically in any linear transformation, see Section~\ref{sec:distortion} and \citet{Fritz_09}. 
However, the strength of the refraction also depends on the wavelength of the light. While that dependence is small in the near-infrared \citep{Fritz_09}, the effect is larger in the optical \citep{Kaczmarczik_09}.
In the optical:
\begin{equation}
[n(\lambda)-1]\cdot 10^6=64.328+\frac{29498.1}{146-(1/\lambda)^2}+\frac{255.4}{41-(1/\lambda)^2}
\end{equation}
is valid \citep{Filippenko_82}.
While the absolute refraction does not matter here, DCR 
causes a wavelength-dependent angular offset, $\beta_\mathrm{offset}$, between objects of different wavelengths. To correct for $\beta_\mathrm{offset}$, it is necessary to know the effective wavelength ($\lambda_\mathrm{eff}$) of an object within a given bandpass. For photon counting detectors
the definition is
\citep{Fritz_11,Tokunaga_05}:   
\begin{equation}
\lambda_\mathrm{eff}=\frac{\int \lambda^2 F_\lambda (\lambda) S(\lambda) }{\int \lambda F_\lambda (\lambda) S(\lambda) }
\end{equation}
Therein $S(\lambda)$ is the full transmission curve of the filter used, including the atmospheric transmission\footnote{see http://classic.sdss.org/dr7/instruments/imager/ for SDSS.},
 and $F_\lambda (\lambda)$ is a flux-calibrated spectrum.
Thus, flux-calibrated spectra are necessary for the calculation of effective wavelengths. 
From these spectra we derive for each object class the color-$\lambda_\mathrm{eff}$ relation. These relations can be converted into color-$\beta_\mathrm{offset}$ relations for a given $\zeta$, see Figure~\ref{fig:dcr1}. SDSS uses flux-calibrated star spectra from \citet{Gunn_83}, see \citet{Pier_03}. 
We also use these spectra for our stars to derive the color-$\lambda_\mathrm{eff}$ relation. To construct a color for our sources, we use the $i$-band. The $i$-band is the best band to derive a color-$\lambda_\mathrm{eff}$ relation since it is closest in $\log(\lambda)$ to the $r$-band. It also contains less strong features than the $g$-band.

We next test whether galaxies follow the same color-$\lambda_\mathrm{eff}$ relation as the stars.
Since SDSS spectra do not cover the full imaging magnitude range (down to $r=21.5$), we instead use spectra from the zCOSMOS bright sample\footnote{Based on observations made with ESO Telescopes at the La Silla Paranal Observatory under program ID 175.A-0839.} \citep{Lilly_09} to obtain $\lambda_\mathrm{eff}$ (and both datasets in the calculation of colors, as described below).
Compared to some other public samples this sample has the advantage that the spectra also cover the $i$-band. 
The two data sets (SDSS and zCOSMOS) differ in their treatment of Galactic extinction.  The Galactic extinction is calibrated out in the stellar spectra while it is present in the  galaxy spectra. In principle, it would be best to use extinction-free spectra in both cases, and to use these to also calculate the effect of extinction.
However, we deem that this is unnecessary in our case: the extinction to Pal~5 (A$_r-$A$_i=0.026$ to 0.048) and towards the zCOSMOS field (A$_r-$A$_i=0.011$ to 0.016) is small compared to the 
color range of stars ($r-i \sim-0.4$ to 1) and galaxies $(r-i \sim 0.1$ to 1.2).
 Furthermore, the extinction-related DCR correction follows approximately the same function of color as the intrinsic color-$\beta_\mathrm{offset}$ correction.

For stars we derive a tight color-$\beta_\mathrm{offset}$ 
relation, see Figure~\ref{fig:dcr1} (we exclude some noisy very late-type dwarfs).
However, a linear relation, as used by SDSS \citep{Pier_03}, does not fit the full color range. We instead use a spline relation. For normal-colored stars the spline is close to the linear relation used by SDSS. We set $\beta_\mathrm{offset}$ of stars with $r-i=0$ to zero.

For the galaxy relation we use 303 galaxies from the zCOSMOS sample to calculate $\lambda_\mathrm{eff}$. This is a  sufficient number to not be limited by statistics.
The high S/N 
part of these spectra lies redward of 0.562 $\mu$m. To obtain an estimate of the spectra blueward of this out to 0.538 $\mu$m (the blue edge of the $r$-band filter), we fit the range between 0.562 and 0.577 $\mu$m and extrapolate.  We note that this is only a small extrapolation, as this range contains only a small fraction of the total $r$-band flux.

We then quantitatively examine the resulting $(r-i)$-$\lambda_\mathrm{eff}$ relation for the galaxies, as a function of both the SDSS colors and the synthetic colors calculated from the spectra. The $(r-i)$-$\lambda_\mathrm{eff}$ relation has more scatter and contains more outliers when we use the SDSS colors. That is to be expected, since the SDSS data is noisier. We therefore do not use SDSS colors of individual galaxies for the determination
of the $(r-i)$-$\lambda_\mathrm{eff}$ relation. The median color difference between the synthetic and the SDSS galaxy colors is 
 $(r-i)_\mathrm{syn}-(r-i)_\mathrm{SDSS}=0.092\pm 0.009$. 
This offset could be physical, because SDSS measures the full galaxy color while the zCOSMOS slits cover only the (usually redder) galaxy centers. However, the offset is  more likely a result of a spectral calibration issue. We explore the astrometric effects of both possibilities 
below.

We exclude 17 galaxies from our sample, due to noise or extreme emission lines. We examine the resulting $(r-i)$-$\beta_\mathrm{offset}$ relations in Figure~\ref{fig:dcr1}. 
The $(r-i)$-$\beta_\mathrm{offset}$ relation for galaxies is noisier than for stars. This is expected from the fact that galaxies have more variable extinction. The galaxy $(r-i)-\beta_\mathrm{offset}$ relation is also offset from that of stars. This offset is caused mainly by the fact that galaxies consist of many types of stars. To test this we tested how extinction-free, single stellar populations, such as those of \citet{Coelho_07}, look on such a diagram, and found that a single-population sequence is itself offset from our stellar sequence by about the same amount as the galaxy spectra.

We fit the galaxy  $(r-i)-\beta_\mathrm{offset}$ relation with a linear model. A fourth order polynomial reduces the scatter by only 6\%, and we therefore stay with a simple linear model.
We set the few very red ($r-i>1.1$) and blue ($r-i<0.1$) sources to have a constant offset given by the value at the border of their defining range, and fit the remaining galaxy data with a linear model, obtaining
 \begin{equation} \beta_\mathrm{offset}=(11.17\pm0.24) -(35.76\pm1.00)\times (r-i)\, \mathrm{[mas]}
 \label{eq:dcr_cor_gal}
 \end{equation} 
 for $\zeta=45\degree$.
 The difference between model and data has a scatter of 3.9 mas.

We then test how a spectral 
 calibration issue would affect the astrometry. In this case we assume that the correct synthetic flux, $F_\mathrm{cal}$, can be obtained in the following way from input spectra, $F_\mathrm{input}$:
\begin{equation}
F_\mathrm{cal}=F_\mathrm{input}\times\lambda^x 
\label{eq:cor_lam}
\end{equation} 
From fitting the above equation with the given flux inputs, we get $x=-0.5$. 
For the color term in Equation 5 we use the median for galaxies in the area of our Pal~5 data, which is $r-i=0.48$.
 For this median color, $\beta_\mathrm{offset}$ changes by only 0.3 mas when we use  $F_\mathrm{cal}$  instead of $F_\mathrm{input}$ from Equation~\ref{eq:cor_lam}. This is comparable to the 1-$\sigma$ error of Equation~\ref{eq:dcr_cor_gal}. 
 The change is small because the changes in $r-i$ and $\beta_\mathrm{offset}$ work to cancel each other out. 
Because the difference is not significant, it does not matter for the astrometric accuracy whether the spectra are offset due to a calibration issue or
 by a physical effect.
 The fact that our DCR zero-point is anchored by stars and not an `average galaxy'  has the consequence that the positions of our galaxies are offset on average. However, we are not interested in perfect absolute positions for galaxies, we only want a reference system which does not move due to DCR, which we achieve with the correction from Equation~\ref{eq:dcr_cor_gal}.

For the application of the correction we first divide the sources into stars and galaxies as in Section~\ref{sec:selection}. Note that quasars are stars according to this classification. Since their spectra are different from both stars and galaxies \citep{Kaczmarczik_09} our DCR correction will likely not work on them. Thus, we do not use them for tests and ignore them in this work. We also need to use a different DCR treatment for SDSS versus LBC positions. The SDSS positions are already in equatorial coordinates and are corrected for DCR. Since we have derived here a better DCR correction we subtract the applied DCR correction from the SDSS positions and then apply our new one.
In the case of LBC, we derive a linear coordinate transformation from pixel to equatorial coordinates for each image and detector, to convert  $\beta_\mathrm{offset}$ from mas to pixels. 
 Our transformation assumes perfect alignment of the pixel axes with R.A. and Dec. While this is not exactly true, the deviations from these assumptions are smaller than 2\% in total. We then apply the pixel DCR corrections to the pixel positions. 
 The corrected pixel positions are then converted back to equatorial coordinates (see Section~\ref{sec:distortion}).

 The mean difference between the raw and the DCR-corrected motions is  $\delta\mu_\alpha=-0.03$ mas/yr and $\delta\mu_\delta=0.38  $  mas/yr.  
 From the 3.9 mas scatter, and factoring in the number of galaxies and the time baseline, we derive an error of less than 0.02 mas/yr for the final DCR correction.
 Even if this error is underestimated due to unknown systematics it is not significant compared to our total proper motion error of about 0.18 mas/yr per dimension.

\subsection{Reference frame and target star selection}
\label{sec:selection}

To obtain proper motions it is necessary to identify member stars and to measure their motions relative to a reference system with a known velocity. In principle, point sources are preferable as a reference since the position of sources with small FWHM  can be measured more precisely \citep{Fritz_09}. However, there are no good point sources in our case. The density of quasars is too low for sufficient precision.
The average velocity of many faint halo stars is larger than our measurement accuracy (see e.g.~\citet{Bond_10}).
Thus, stars can only be used when their velocities are known with high precision. There is no reliable option in our case: the Hipparcos astrometry catalog \citep{Hog_00} only contains stars that are saturated in our data and their density is also too low.
Photographic plate-based survey catalogs such as those of \citet{Munn_04} are deeper. However, while they have good per-star precision, it is possible they contain systematic errors. It is risky to use a galaxy model 
(see the example of \citet{Pryor_10} for the Sagittarius dwarf)
to predict the mean velocity of the foreground star population, since Pal~5 is distant from objects with well measured proper motions like Sgr A*.
We thus use galaxies as our reference frame: they are abundant and we also have better control in estimating the measurement errors.

As a starting point for the galaxy sample we use the well vetted morphological classification of SDSS in the r-band. 
A potential worry for the reference galaxy sample is pollution from stars. We reduce the likelihood of this in several ways. 
The likelihood of pollution is especially high towards the dense center of Pal~5. Therefore we check galaxies within $r=0.07\degree$ of the cluster center 
 by eye in the SDSS and LBC data. 
 We find that galaxies which fall approximately along the cluster CMD sequence look star-like and have a star-like FWHM in the LBC images.
 We therefore exclude all galaxies in this CMD region, as well as any other galaxies which appear star-like in shape or FWHM, from our reference sample. 
 
 Furthermore, we use the LBC images to refine our selection. We use SExtractor together with PSFex \citep{Bertin_96, Bertin_11} in the three-step process outlined by \citet{Koposov_15}. We use as the criterion for galaxies {\tt |SPREAD\_MODEL|}$>0.003$ \citep{Desai_12,Annunziatella_13,Koposov_15}. Only 4\% of the galaxies selected according to the previous criteria are rejected by this criterion. We verify by eye that most of these appear star-like in the LBC images. Since all of these criteria need to be fulfilled at once, our galaxy selection  is more robust than when using, say, only the SDSS morphological classification.

 Next we use variants of these criteria to estimate the uncertainty caused by the galaxy selection. The SDSS criterion is  {\tt PSF\,mag-cmodel\,mag}$=\delta_{\rm mag}=0.145$ for each band separately. We test this  criterion in three different ways: first, we use the information from all bands with enough signal ($g$ and $i$-band) in addition to the $r$-band, which is our main workhorse; second,
 we add a significance criterion compared to the magnitude error of galaxies ($\sigma$ {\tt cmodel\,mag}): $\sigma_\mathrm{gal}=(\delta_{\rm mag})/(\sigma$ {\tt cmodel\,mag}); 
 we use $\sigma_\mathrm{gal}=5$, which means we only use objects which are significantly brighter when fit with a galaxy model than when fit with a stellar PSF model. The LBC selection remains the same as described above. 
 In both cases we calculate the change to the final proper motion by the selection, we obtain changes of order $\delta\,\mu_\mathrm{gal}\approx|0.04|$ mas/yr per dimension.
That is small compared to the final total accuracy of 0.18 mas/yr per dimension (Section~\ref{sec:vel}).

\begin{figure}
\begin{center}
\includegraphics[width=0.995 \columnwidth,angle=0]{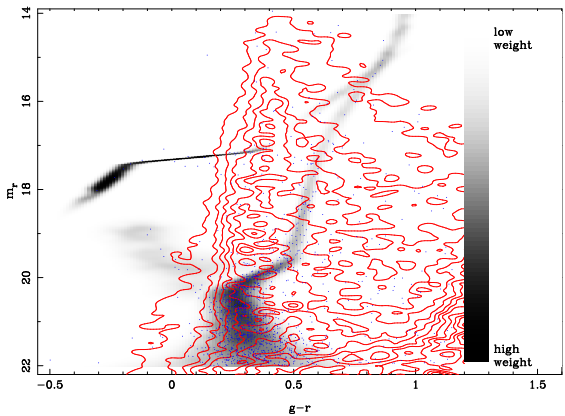}
\caption{CMD of Pal~5 region, using extinction-corrected SDSS {\tt PSF\,mag} values, and excluding very red sources ($g-r>1.2$). 
The blue dots show Pal~5 stars.  The gray background shows the matched filter $W$ which we use to identify Pal~5 stars. The red contour levels in linear steps show the density in the background field.
} 
\label{fig:filter0}
\end{center}
\end{figure}

To identify stars belonging to Pal~5, we use the matched filter technique \citep{Rockosi_02}. The matched filter is defined in the following way:
\begin{equation}
W=\Sigma(CMD)_{\mathrm{target\,object}}/\Sigma(CMD)_{\mathrm{background}}
\end{equation}
$\Sigma(CMD)_{\mathrm{x}}$ above is the density of stars in bins of color and magnitude in a CMD, either the CMD of the target (Pal~5) or the background. For the  CMD we use the $r$ and $g$-bands, which have the highest SNR. We first select all stars in SDSS within 8.4' of the cluster center (the tidal radius of Pal~5 according to \citealt{Rockosi_02}), see Figure~\ref{fig:filter0}. 
We then exclude stars that do not match the majority of Pal~5 stars in the color magnitude diagram via outlier rejection. 

There are several options in constructing the Pal~5 matched filter, $W$.  One option is to do simple binning, as in a Hess diagram.  
However, that is not optimal, because it is influenced by shot noise in the case of a small number of stars. Another option, binning with smoothing, makes the shot noise smaller; however, it also makes the cluster sequences wider than they are in reality. 
 Instead we model the cluster CMD as much as possible:  we fit the main sequence, the (sub)-giant branch and the horizontal branch with different Gaussians whose widths, heights, and positions vary smoothly as a function of magnitude. For the blue straggler region Gaussian modeling does not work well. There we use a smoothed Hess diagram instead. 

 For the background we choose a 4 square degree region that is offset from the Pal~5 cluster, with borders of 225$\degree$ and 227$\degree$ in R.A. and  1.9$\degree$ and 3.9$\degree$ in Dec. This region is large enough that shot noise is not important and small enough that the Milky Way background is still close to identical to the Pal~5 region.  We smooth the Hess diagram of the background slightly to further reduce the influence of shot noise.  The final filter $W$ is shown in Figure~\ref{fig:filter0}. There, and in general, we use a magnitude cut-off of $r=22$ which is approximately the completeness limit of SDSS for point sources. The precise cut-off is not important since stars close to this limit have large motion errors such that they are given negligible weight in the subsequent analysis.

\begin{figure}
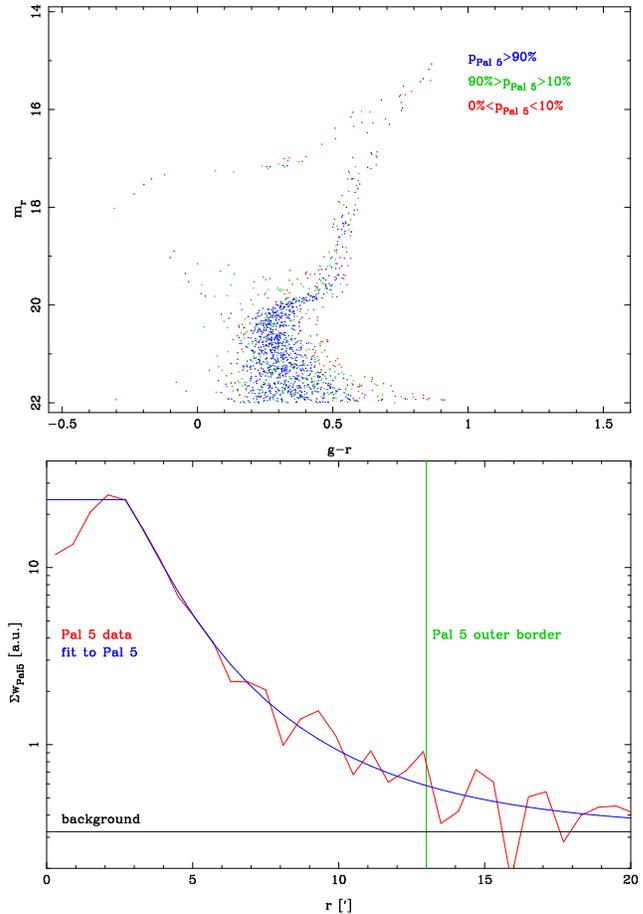

\begin{center}
\includegraphics[width=0.70 \columnwidth,angle=-90]{f4a.eps} 
\includegraphics[width=0.70 \columnwidth,angle=-90]{f4b.eps} 
\caption{Top: CMD of Pal~5 (r$<$13'). The probability for membership depends on the position in the CMD (the value of the matched filter $W$ there, see Figure~\ref{fig:filter0}) and the distance from the cluster center. Stars with zero probability are not plotted. 
Bottom: Radial density profile of Pal~5 after application of the weighted filter. The central depression is caused by source confusion.
We fit the other data with a polynomial.  
} 
\label{fig:filter1}
\end{center}
\end{figure}

To get membership probabilities, $p$, for each star, first a relative weight ($w$) is obtained by checking in the matrix $W$ the value at the magnitude and color of a given star. We then construct the background density ($\Sigma(w)_\mathrm{back}$) and cluster density ($\Sigma(w)_\mathrm{Pal5}$) from these weights. For the cluster, we measure the density radially in bins from the center, see Figure~\ref{fig:filter1}. $\Sigma(w)_\mathrm{Pal5}$ dips in the cluster center because the high source density reduces the number of {\tt clean} SDSS sources. 
We mask out this central depression and fit the density of the other radial bins with a polynomial. 
To measure the density of the background we use a 2$\degree$ box, centered on the cluster, from which we mask out the cluster and the stream. The radial membership fraction is defined as $f_\mathrm{Pal5}(r)=(\Sigma(w)_\mathrm{Pal5}-\Sigma(w)_\mathrm{back})/\Sigma(w)_\mathrm{Pal5}$.
We then obtain a probability $p$ for every star using the following procedure:
for each $f_\mathrm{Pal5}(r)$ we solve the following equation for $w_\mathrm{lim}$: 
\begin{equation}
f_\mathrm{Pal5}(r)=\frac{\int_{w_\mathrm{lim}}^\infty W}{\int_0^\infty W}
\label{eq:f_r}
\end{equation}
 We then draw for each star a random number (between 0 and 1) and multiply it by its $w$. 
When that product is larger than $w_\mathrm{lim}$ the star is used. This way we create 10 different random samples, in order to test whether the uncertainties in membership are important.
We do not use stars outside of $r=13'$. At $13'$ $f= 45$\%.

The error contribution to the proper motion from this Pal~5 star selection, as calculated from the scatter of the 10 different samples, is small. We obtain a contribution of only 0.046 mas/yr. 
This is less than the contribution from distortion and image registration (see Section~\ref{sec:distortion}).
We did not test whether galaxies misclassified as stars are a relevant problem. However, because we select Pal~5 stars in the CMD and because there are more Pal~5 stars than galaxies in the cluster region, this effect should be small.

\subsection{Distortion correction and image registration}
\label{sec:distortion}

Distortion is often the main error source in astrometry \citep{Fritz_09}.
There are two main ways to estimate the distortion. Firstly, one can use many dithered observations to infer the distortion field of a given instrument without external reference (see e.g. \citealt{Anderson_03,Trippe_08,Bellini_10}). 
Secondly, one can use a distortion-free source list of the observed area. 
 While the first strategy has  higher position precision (a position error of $\sigma X_{1D}=10$ mas in the case of LBC, see \citealt{Bellini_10}) 
it is necessary to be able to apply it to all epochs. This is a disadvantage for us because we use different instruments. While such a correction would be possible for the LBC, it is more difficult in the case of SDSS. It is further unclear whether improvements can be made over the existing SDSS distortion calibration from \citet{Pier_03}.

For our correction we use the SDSS astrometry as reference. According to \citet{Pier_03} the SDSS internal precision is about 25 mas.
 Thus, we expect a similar accuracy. In addition to the distortion, it is also necessary to solve for the image registration (linear terms). In our definition the image registration contains the linear parameters and the equatorial coordinate of the central pixel, which changes according to the pointing. 
Unlike the distortion terms which can be stable over time, the linear parameters  are usually not stable \citep{Bellini_10}, due to different effects, for example, the changing airmass. Constant image scale parameters limit the accuracy to about 1$\arcsec$ \citep{Bellini_10}.
 Thus, it is necessary to solve for all image registration parameters in every image.

In practice we use the following process: first, we apply the distortion correction to the raw pixels to obtain undistorted pixel coordinates\footnote{Before that we exchange  x and y definitions for detector 4. Then they are approximately aligned with the others detectors.}. Second, we apply the image registration to the corrected pixels to obtain sky coordinates. 
 For the distortion correction we use the following equation, wherein $x'$ and $y'$ are measured relative to pixel (1049, 2304) respectively, the approximate center of the detectors:
 
 \begin{equation*}
 \begin {split}
x_\mathrm{cor}=x+a_1\,x'^2+a_2\,x'y'+a_3\,y'^2+a_4\,x'^3\\+a_5\,x'^2y'+a_6\,x'y'^2+a_7\,y'^3
\end{split}
\end{equation*}

\begin{equation}
 \begin {split}
y_\mathrm{cor}=y+b_1\,x'^2+b_2\,x'y'+b_3\,y'^2+b_4\,x'^3\\+b_5\,x'^2y'+b_6\,x'y'^2+b_7\,y'^3
\end{split}
\label{eq:dist_corr}
\end{equation}

Using higher order corrections than this do not improve the residuals. 
 Our residual scatter for bright galaxies is close to the internal precision of SDSS. 
The distortion correction does not contain linear terms because these are contained in our image registration transformation:

 \begin{equation*}
\mathrm{R.A.}=c_1+c_2\,x_\mathrm{cor}+c_3\,y_\mathrm{cor}
\end{equation*}

\begin{equation}
\mathrm{Dec.}=d_1+d_2\,x_\mathrm{cor}+d_3\,y_\mathrm{cor}
\label{eq:lin_corr}
\end{equation}

We test three different variants of pixel coordinate transformation: firstly, time-dependent (separate for each image) and detector-dependent distortion parameters and image registration which, as stated, vary with time and detector.  Secondly,  time-independent but detector-dependent distortion and image-registration parameters. Thirdly, time-independent but detector-dependent distortion parameters and time-dependent image registration, which are coupled between the different detectors. Specifically, in the third case, we first fit the same time-dependent image registration parameters to all detectors. We then fit a constant skew and shift term for  each detector except for detector one which we use as the ``master'' detector. 
This case has clearly larger residuals than the other two cases, confirming the finding of \citet{Bellini_10} that using all LBC detectors together leads to larger errors than when the detectors are treated separately.
In a variant of case three we couple the offsets between the detectors to the variable image scale. This assumes that only the airmass changes the angular distances between the detectors.
In this variant the residuals are reduced but are still larger than when the detectors are treated independently.  
Thus the detector distances depend on at least one other variable, possibly  temperature variations.
We do not use case three because of its larger errors.

The first two cases give consistent results; that means the distortion is stable over the 80 minute time-scale of our observations. 
 We decide to use case two: independent detectors with a constant distortion solution for each, but time-variable image registration terms. Compared to case one it has the advantage that it has fewer free parameters.
 
 The small number of galaxies in the very core of Pal~5 (see Figure~\ref{fig:datdis}) affects the precision of the linear parameters adversely.
We therefore use the cluster stars  as additional reference sources.
Since the cluster stars move, we add the displacement in R.A. and Dec as additional parameters. 
These two parameters can be easily converted to the proper motion of Pal~5, by dividing through the time baseline between the two epochs. This is easily possible because the time baseline is effectively the same for all stars, since there is only a one day difference in the observing epochs of the Pal~5 cluster in SDSS, and only a few minute difference in LBC.

Adding the distortion terms and the image registration terms together we fit 334 parameters for each of the four detectors. The number of sources used in the fit depends on the detector and the drawing of Pal~5 stars. There are about 11,970 sources per detector. There are always more than enough sources to fit the parameters.
For the fitting we use {\tt mpfit} \citep{Markwardt_09}. For the position errors of the objects we use the SDSS errors ($\sigma X_\mathrm{1D\,SDSS}$, called {\tt raErr} and {\tt decErr} in SDSS) as a base. 
From that we calculate the full error:
\begin{equation}\sigma X_\mathrm{1D}=\sqrt{(\sigma X_\mathrm{1D\,SDSS}\times fact)^2+add^2}
\label{eq:err_model}
\end{equation}
$add$ is the systematic, magnitude-independent error, which is not included in the $\sigma X_\mathrm{SDSS\,1D}$ of DR9.
$fact$ can stand for LBT-related errors and/or serve to correct  $\sigma X_\mathrm{SDSS\,1D}$.
As a first guess we use $add= 20$ mas and $fact=$1.   
In our first derivation of the transformation terms we fit only galaxies. We calculate the residual ($R$) of the positions relative to the fit, and exclude outliers with $R=\sqrt{R_x^2+R_y^2}>10\,\sigma$. A possible explanation for outliers are objects misclassified in SDSS, bad pixels, and in the case of complex galaxy morphologies, centering problems.  

\begin{figure}
\begin{center}
\includegraphics[width=0.70 \columnwidth,angle=-90]{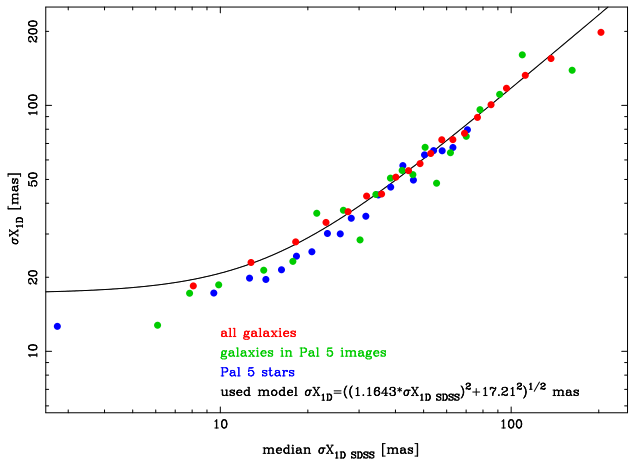} 
\caption{Residual position scatter of LBC sources compared to SDSS sources. We use $1.483\times$ the median deviation as a robust scatter measure. The line shows our error model.
} 
\label{fig:residuals1}
\end{center}
\end{figure}

We then analyze the residuals (after outlier rejection) in the following way: we bin these galaxies according to $\sigma X_\mathrm{1D\,SDSS}$ in 20 equally-populated bins, see Figure~\ref{fig:residuals1}. In each bin we determine the scatter of the residuals between the two epochs. For robust outlier-rejection we do not use the standard deviation, but the median deviation. We scale it by a factor of 1.483, to make its value identical with the standard deviation under the assumption of a Gaussian distribution. 
We fit the error model (Equation~\ref{eq:err_model}) to the points, finding that 
it is a good representation of our data as can be seen in Figure~\ref{fig:residuals1}. Only the point for the largest errors is clearly  offset, possibly due to the fact that the error for the point may be overestimated. 
We exclude it; due to its large error, its influence is very small.

We repeat the fit several times with iterative outlier rejection. In the process we also add the Pal~5 stars to the data sample. We use for them the same error model, since the error should not depend on the nature of the source.  The parameters of our final error model are  $add=$17.21 and $fact=1.186$, see Figure~\ref{fig:residuals1}.
It seems slightly conservative especially in $add$ for the objects in the Pal~5 cluster center field. However, it is difficult to measure $add$ reliably from this relatively small area. 
For the final outlier rejection we choose $R=3.21$. 
With this cut the final $\chi^2/d.o.f.=1$, which implies that our sample does not contain further outliers. The distribution is not fully Gaussian; there are a few more stars with very small errors and relatively large residuals than expected from a Gaussian distribution.
As a test we also model the errors by fitting only cluster stars or only galaxies, finding that the obtained motions do depend slightly on the error model used. From the scatter in these different results, we obtain a motion error of 0.04 mas/yr, which we fold into our final errors below.

\begin{figure*}
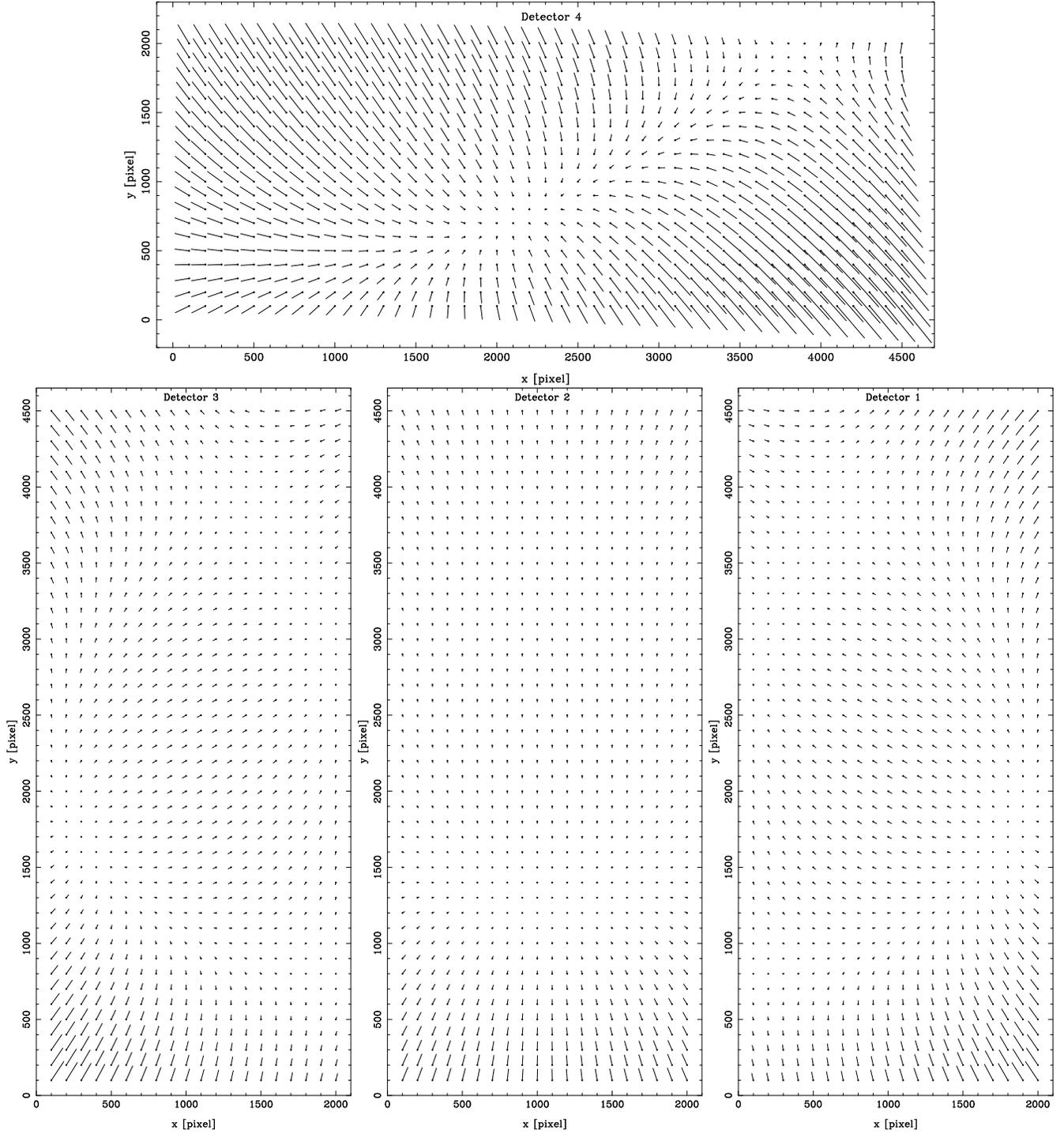

\begin{center}
\includegraphics[width=0.75 \columnwidth,angle=-90]{f6a.eps} 
\includegraphics[width=1.45 \columnwidth,angle=-90]{f6b.eps} 
\includegraphics[width=1.45 \columnwidth,angle=-90]{f6c.eps} 
\includegraphics[width=1.45 \columnwidth,angle=-90]{f6d.eps} 
\caption{Distortion field of LBC. We show here for each of  the four detectors the distortion field as a function of the original pixel positions.
The vectors show the difference between the corrected and original pixel positions enlarged by a factor of 4. We label the detectors as in the header naming convention. 
} 
\label{fig:dist_field}
\end{center}
\end{figure*}

Overall we have enough coverage (Figure~\ref{fig:datdis}) to obtain a robust distortion solution, see Figure~\ref{fig:dist_field}.
The distortion is qualitatively similar to the one obtained by \citet{Bellini_10} and \citet{Fabrizio_14}. The differences can be explained by the use of a different filter and camera and by time variability on the timescale of years.
Our solution has a 1D accuracy of  17 mas over the full area. (Using only the cluster the accuracy is 13 mas.)
\citet{Fabrizio_14} obtain $\sigma X_\mathrm{1D}=6.7$ mas as precision for LBC. The difference is likely caused by the fact that our error estimate contains also the SDSS errors, which alone are the size of our $\sigma X_\mathrm{1D}$ \citep{Pier_03}.

\subsection{Proper motion of Pal~5}
\label{sec:vel}

To obtain the final motion we combine the Pal~5 displacements of the four different detectors weighted by their errors. 
For the final errors, we first correct for the fact that the displacements obtained from the different LBC images are correlated. The reason is that they all use the same SDSS positions, which dominate the errors.  We first multiply the errors by $\sqrt{N}$, where $N=10.5$, is the average number of detected galaxies. This assumes that only SDSS contributes to the error budget. To include non-zero LBC errors, we need to multiply this by another factor, $M\leq1$.
From the strength of the position correlations between the different LBC images we derive $M=0.944$.
Combining the two factors,  we derive $\sigma\mu_{\alpha}=0.176$ mas and  $\sigma\mu_{\delta}=0.172$ mas from the galaxy and star errors ($\sigma X_\mathrm{1D}$).
To this error term we also add the uncertainties from image distortion and registration (Section~\ref{sec:distortion}), and the uncertainty due to the Pal~5 membership selection (Section~\ref{sec:selection}). Other error sources contribute negligibly. 
Considering all these effects we obtain  $\mu_{\alpha}=-2.296\pm0.186 $ mas/yr and $\mu_{\delta}=-2.257\pm0.181$ mas/yr. 

\begin{figure}
\begin{center}
\includegraphics[width=0.99 \columnwidth,angle=-90]{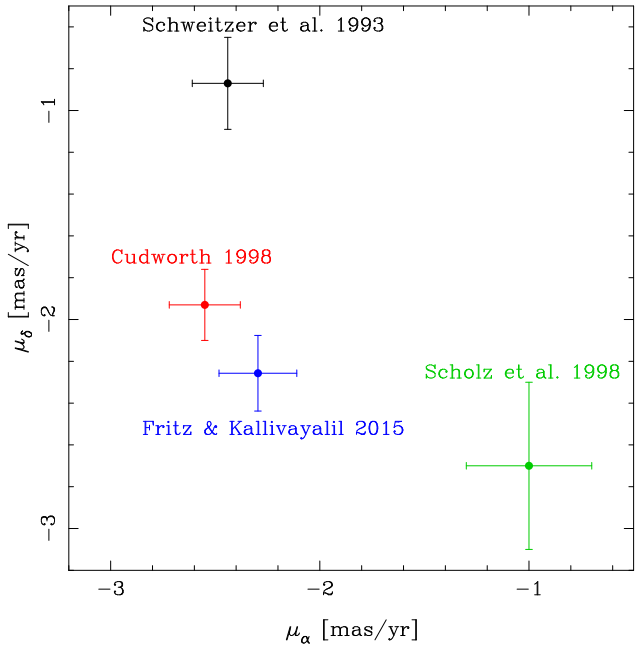} 
\caption{Proper motion measurements of Pal~5.
} 
\label{fig:mot_lit}
\end{center}
\end{figure}

In Figure~\ref{fig:mot_lit} we compare our measurement (the first to be obtained from CCD data) with the values from photographic plate studies in the literature  (\citealt{Schweitzer_93,Scholz_98}, K. M. Cudworth 1998, unpublished, see \citealt{Dinescu_99}). The scatter between the photographic plate-based motions is larger  than their quoted errors. The mean of these measurements 
is roughly consistent 
with our value. 

 \section{Constraints on the Milky Way Halo}
\label{sec:MW_mass}

We now use the 6D phase space properties of Pal~5, i.e., its position, distance, radial velocity, and measured proper motion, and the publicly available code \textit{galpy}  
\citep{Bovy_14b},
to model the disruption of the Pal~5 system.
 We compare the predictions of the model to the measured location of the stream debris and the observed radial velocity gradient along the stream.

\subsection{Used parameters}
\label{sec:parameters}

While the position of Pal~5 is known to sufficient precision, its distance is more uncertain. 
\citet{Harris_96} derives a distance of 23.2 kpc from the horizontal branch, as is used by \citet{Odenkirchen_09}.
\citet{Vivas_06} derive 22.3 kpc from RRLyrae.
\citet{Dotter_11} obtain 20.9 kpc\footnote{Note that the distance moduli cited in \citet{Dotter_11} are not corrected for extinction (private communication), and therefore need to be corrected before use.}
 from fitting an HST CMD. 
It is unclear from the respective error analyses whether these values are in agreement with each other.
In order to explore a realistic range, we usually use the smallest (20.9 kpc) and largest (23.2 kpc) values as a bound. As our ``standard'' value we use the mean of these values: 22.05 kpc.

The radial velocity of the cluster, $V_r$, is -58.7 km/s \citep{Odenkirchen_02}. A comparison with \citet{Kuzma_14} shows that the error is approximately $0.8$ km/s. The measured dispersion of the cluster is $\sigma_\mathrm{Pal5}(rad)=1.1\pm0.2$ km/s, which upon correction for inflation caused by binary stars is $\sigma_\mathrm{Pal5}(rad)\approx0.3\pm0.15$ km/s \citep{Odenkirchen_02}. 
Due to mass loss, the dispersion of Pal 5 was higher in the past \citep{Dehnen_04}. Therefore, we use a value of $\sigma_\mathrm{Pal5}(rad)= 0.4$ km/s in our modeling below.

For the stream we use two observables, the stream streak on the sky and the radial velocity gradient along the stream.
The radial velocity gradient along the stream was measured by \citet{Odenkirchen_09} and \citet{Kuzma_14}.
\citet{Odenkirchen_09} measured velocities for Pal~5 stars using high resolution spectroscopy. In their work they use only spectroscopically confirmed giants. Four giants are clearly not associated with Pal 5 because of their radial velocities. In the case of two others it is less clear whether they are members, binaries whose velocity is influenced by their partner, or are not members.

\citet{Kuzma_14} use (mostly) medium resolution spectroscopy to obtain giants over a larger range of the stream. For selecting the Pal~5 stars they use spectroscopic giant-dwarf and metallicity discriminators. Finally, they exclude stars above -30 km/s and below -70 km/s. 
The stars at the largest distances from the cluster are somewhat off the stream location (as measured by us and \citet{Grillmair_06}. While this maybe worrisome, their simulation shows that pollution by non-stream stars should not be a relevant problem. Therefore, we use all the members of \citet{Odenkirchen_09} and \citet{Kuzma_14} together with the cluster velocity to obtain the gradient as a function of R.A.  Since we are interested in the mean velocity of the stream, we add in quadrature to the individual measurement errors the stream dispersion of 2.1 km/s 
\citep{Kuzma_14}. First, we also try to include the two dubious members from \citet{Odenkirchen_09}, but that results in $\chi^2/\mathrm{d.o.f.}=116.2/56$. Excluding them, we obtain $\chi^2/\mathrm{d.o.f.}=47.4/54$. Thus, the velocities of these two stars are very likely influenced by some non-stream motion (such as binarity). We therefore exclude them and obtain a gradient of $1.21\pm0.09\,\mathrm{km}\,\mathrm{s}^{-1}\mathrm{deg}^{-1}$ in R.A. The gradient is identical within the errors when using all stars, or only the stars from either \citet{Odenkirchen_09} or \citet{Kuzma_14}.
Most stars at large distances from the cluster are on the same side of the stream, which causes some additional uncertainty. 
To account for this as well as other potential uncertainties,
we enlarge the errors slightly and use $1.20\pm0.15\,\mathrm{km}\,\mathrm{s}^{-1}\mathrm{deg}^{-1}$ as gradient when we compare models with the gradient.

\begin{figure}
\begin{center}
\includegraphics[width=0.70 \columnwidth,angle=-90]{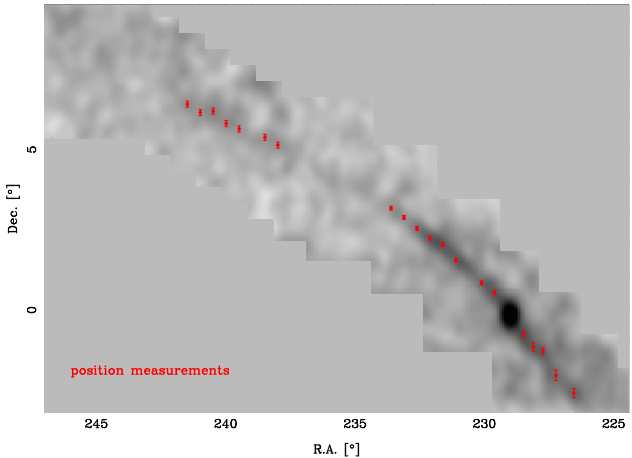} 
\caption{Matched filter map of the Pal~5 stream. The red points show the stream positions which we use in our modeling.
} 
\label{fig:stream_streak}
\end{center}
\end{figure}

As for the stream streak, because it is difficult to estimate the errors from published maps, 
we construct our own.  
We retrieve all stars from SDSS for the stream and adjacent areas.
We begin by binning the stars in quadratic bins of 3.75' in length. 
For map one we give all the stars the same weight. For map two we weight the stars with the filter constructed in Section~\ref{sec:selection}.
To account for varying surface density along the stream, we divide map two by map one. We treat bins with less than 20 stars as bad pixels and interpolate over them. 
 To increase the signal to noise ratio of this map we smooth it with a 30' Gaussian, obtaining the final map of Figure~\ref{fig:stream_streak}.
We sample the map at 30' R.A. intervals, and fit a Gaussian plus a gradient at each location. As a first-guess error for each position, we ensure that the relative errors account for the variable density of the stream. We treat the leading and trailing streams separately, since the stream follows a complex shape (an `S') close to the cluster. A linear model is sufficient for the short leading stream, while we use a quadratic model
for the longer trailing stream. 
Finally, we rescale the previous errors by a common factor to achieve  $\chi^2/d.o.f.=1$ for the leading and trailing streams separately.
We present the obtained positions in  Table~\ref{tab:_par1}. Our positions agree well with the recently published positions of \citet{Kuepper_15}

\begin{deluxetable}{lll} 
\tabletypesize{\scriptsize}
\tablecolumns{3}
\tablewidth{0pc}
\tablecaption{stream streak positions \label{tab:_par1}}
\tablehead{ R.A. [$\degree$] &  Dec. [$\degree$] &   $\sigma$Dec. [$\degree$]}
\startdata
241.48 & 6.41 & 0.09 \\ 
240.98 & 6.15 & 0.09 \\ 
240.48 & 6.20 & 0.09  \\ 
239.98 & 5.81 & 0.09 \\ 
239.48 & 5.64 & 0.09  \\ 
238.48 & 5.38 & 0.09 \\ 
237.98 & 5.14 & 0.09 \\  
233.61 & 3.17 & 0.06  \\ 
233.11 & 2.88 & 0.06  \\ 
232.61 & 2.54 & 0.06 \\ 
232.11 & 2.23 & 0.06  \\ 
231.61 & 2.04 & 0.06 \\ 
231.11 & 1.56 & 0.06  \\ 
230.11 & 0.85  & 0.06 \\ 
229.61 & 0.54  & 0.06  \\ 
228.48 & -0.77 & 0.11  \\ 
228.11 & -1.16 & 0.14  \\  
227.73 & -1.28 & 0.11  \\ 
227.23 & -2.03 & 0.17 \\  
226.55 & -2.59 &  0.14   \\  
\enddata
\end{deluxetable}

The velocity and position of the sun are also necessary inputs. 
For the distance $R_0$ of the sun to the  Galactic center we combine three recent high-accuracy measurements:
\citet{Dekany_13} obtained $8.33\pm0.15$ from RR Lyrae in the bulge, \citet{Reid_14} obtained $8.34\pm0.14$ kpc from parallaxes and velocities of masers in 
a substantial part of the disk, and \citet{Chatzopoulos_14} obtained  $8.27\pm0.13$ kpc from a nuclear cluster model fit to radial velocities and proper motions from \citet{Fritz_14}. The stated errors combine the statistical and systematic errors.  Combining these consistent measurements weighted by their errors we obtain $R_0=8.31\pm0.08$ kpc. 
This $R_0$ is also consistent with most older measurements \citep{Genzel_10}, but has a smaller error.
We also use the following value for the distance $z$ of the sun relative to the mid plane, which is $z=0.02 \pm 0.007$ kpc \citep{Joshi_07,Majaess_09,Buckner_14}.

It is standard to dissect the solar motion into the circular rotation around the Galactic Center, $V_0$, 
 and its peculiar motion, $V_{\mathrm{LSR}}$ \citep{Kerr_86}.
Since $V_0$ and $V_\mathrm{LSR}$ are both uncertain at possibly the 20 km/s level \citep{Schoenrich_10,Bovy_12b,Reid_14},
a better approach for our purposes is to use the motion of the sun more directly. The proper motion of the Sun is visible as reflex motion in the proper motion of Sgr~A* \citep{Reid_04}. Combining the proper motion with the distance gives $V_\mathrm{Sun}(\phi)$ and  $V_\mathrm{Sun}(z)$.
This $V_\mathrm{Sun}\,(z)$ is consistent, but less precise, than the $V_\mathrm{Sun}\,(z)$ determined from the LSR. We therefore use the $V_\mathrm{Sun}(\phi)$ from above, and the \citet{Schoenrich_10} values for  $V_\mathrm{Sun}\,(z)$
 and  $V_\mathrm{Sun}\,(r)$.
 These solar motion components are not disputed \citep{Schoenrich_10,Bovy_12b}.
Our final solar values are: $V_\mathrm{Sun}(r,\phi,z)=-11.1\pm0.7, 251.3\pm2.6, 7.3 \pm0.4$. 
 (Due to our coordinate system definition $V(r)$ is reversed in comparison to $U$ in the LSR.)
Because of our derivation method,  $V_\mathrm{Sun}(\phi)$ and $R_0$ have a strong positive correlation.
 These and all other parameters for the Pal~5 system (apart from the stream streak) are summarized in Table~\ref{tab:_par2}.

\begin{deluxetable}{ll} 
\tabletypesize{\scriptsize}
\tablecolumns{2}
\tablewidth{0pc}
\tablecaption{Parameters of Pal~5 cluster, stream and Sun\label{tab:_par2}}
\tablehead{ parameter &  value }
\startdata
R.A. & 229.018$\degree$ \\ 
Dec. &  -0.124$\degree$ \\
distance & 20.9 to 23.2 kpc \\ 
$\mu_{\alpha\mathrm{Pal5}}$ & $-2.296 \pm 0.186 $ mas/yr\\
$\mu_{\delta\mathrm{Pal5}}$ & $-2.257 \pm 0.181 $ mas/yr\\
$v_\mathrm{Pal5}(rad)$ & $-58.7\pm0.8$ km/s\\
$\sigma_\mathrm{Pal5}(rad)$ & 0.4 km/s\\ 
$v_\mathrm{stream}(rad)$ gradient (R.A.) & $1.2\pm0.15\,\mathrm{km}\,\mathrm{s}^{-1}\mathrm{deg}^{-1}$\\
$r_\mathrm{Sun}$ & $8.31\pm0.08$ kpc\\
$z_\mathrm{Sun}$ & $0.02\pm0.007$ kpc\\
$V_\mathrm{Sun}(r)$ & $-11.1\pm0.7$ km/s\\
$V_\mathrm{Sun}(\phi)$ & $251.3\pm2.6$ km/s\\
$V_\mathrm{Sun}(z)$ & $7.3\pm0.4$ km/s\\

\enddata
\end{deluxetable}

\subsection{Galaxy model setup}
\label{sec:gal_model}

Thanks to the mass-ratios involved, $m_{\mathrm{sat}}<< M_{\mathrm{host}}$, satellite debris streams provide strong constraints on the host potential, and for computational simplicity, streams are often modeled as generally following orbits.
 However, streams do not strictly follow orbits, see e.g. \citet{Johnston_98,Helmi_99,Dehnen_04,Kuepper_12,Sanders_13,Bovy_14}. N-body simulations have shown that stream-orbit misalignments can be significant, and increase with the mass of the cluster and the eccentricity of the orbit \citep{Lux_13}. While N-body simulations are a conceptually simple method to account for stream-orbit misalignment \citep{Dehnen_04,Law_10}, they are computationally expensive, making it difficult to explore parameter space.  There are now several alternative theoretical approaches to obtaining stream positions and velocities, e.g. \citet{Sanders_14a, Bovy_14} 
who use an action-angle formalism, and \citet{Varghese_11,Lane_12,Kuepper_12,Gibbons_14,Pearson_14,Kuepper_15}, who use a combination of a restricted three-body approach and N-body simulations. 
Here, we use the publicly available action-angle formalism of \citet{Bovy_14}, \textit{galpy} \citep{Bovy_14b}. 

For testing purposes, we use the following Milky Way models. 
Our main model ({\tt Standard}) is very similar to the model used by \citet{Pearson_14}, and to a common parametrization of the Galactic potential \citep{Allen_91}, which uses the IAU recommendation of 220 km/s \citep{Kerr_86}. It has three components: bulge, disk and halo.
For the bulge we use a (spherical) \citet{Hernquist_90} parametrization;  for the disk we use a \citet{Miyamoto_75} parametrization; for the halo we use a spherical logarithmic parametrization.
This and the parameters of the three components are close to identical with the spherical halo model of \citet{Pearson_14}. We list them in Table~\ref{tab:_par_mod}.
Besides this main model we use two variants of it: {\tt Small} and {\tt Massive}. In these variants the mass of bulge and disk are identical to what is in the {\tt Standard} model, but we vary the mass of the halo, similar to the approach of \citet{Kuepper_15}. We choose the mass of the halo such that the rotation curve inside of 30 kpc is in rough agreement with other observations at the solar radius \citep{McMillan_11,Bovy_12b,Penarrubia_14,Reid_14} and out to about 30 kpc \citep{Sakamoto_03,Koposov_10,Kafle_12,Kuepper_15}. We ignore constraints further out, because Pal~5 does not enter these distances.

\begin{figure}
\begin{center}
\includegraphics[width=0.70 \columnwidth,angle=-90]{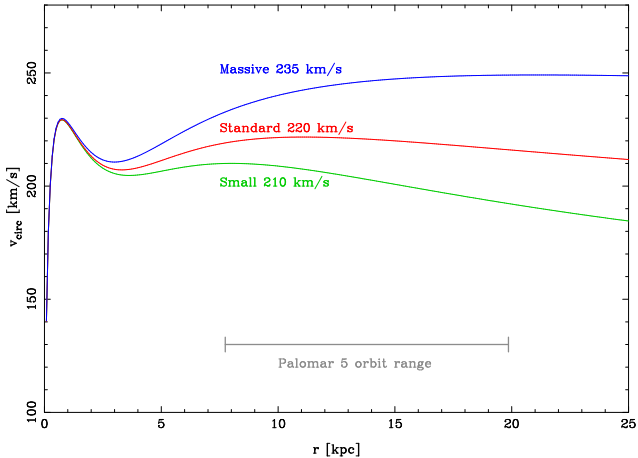} 
\caption{The rotation curves of the three potentials. All use the same disk and bulge, but the (spherical) halo mass is allowed to vary.
} 
\label{fig:rot_vel}
\end{center}
 \end{figure}

A logarithmic potential is probably not a good parametrization of the full halo of the Milky Way. An NFW-profile \citep{Navarro_97,Kuepper_15} is probably a better representation of the Milky Way halo. However, the difference  between a logarithmic and an NFW halo is small over the limited radial range probed by Pal~5. The chosen models have V$_0$  in the range of 210 km/s to 235 km/s. A better measure to characterize the models in our context is the velocity at the apocenter of Pal~5, see \citet{Kuepper_15}. For simplicity we assume here that the apocenter is at 20 kpc in the disk plane.  The velocity there (V$_\mathrm{20}$)
is between 192 km/s and 249 km/s, while our {\tt Standard} model has a value of 218 km/s. The rotation curves of the three models are shown in Figure~\ref{fig:rot_vel}.

\begin{deluxetable*}{lllll} 
\tabletypesize{\scriptsize}
\tablecolumns{5}
\tablewidth{0pc}
\tablecaption{Model parameters \label{tab:_par_mod}}
\tablehead{ Model name & Component & $f_i(V_0)$ & $V_{i\,0}$ & parameters}
\startdata
{\tt Standard} & Bulge & 0.307 & 121.9  & $c=0.7$ kpc\\ 
{\tt Standard} & Disk & 0.496 & 154.9 & $a=6.5$ kpc, $b=0.26$ kpc\\ 
{\tt Standard} & Halo & 0.197 & 97.65 & $r_h=12$ kpc \\ 
{\tt Massive} & Bulge & 0.270 & 121.9  & $c=0.7$ kpc\\ 
{\tt Massive} & Disk & 0.4345 & 154.9 & $a=6.5$ kpc, $b=0.26$ kpc\\ 
{\tt Massive} & Halo & 0.2955 & 127.75 & $r_h=12$ kpc \\ 
{\tt Small} & Bulge & 0.337 & 121.9 & $c=0.7$ kpc\\ 
{\tt Small} & Disk & 0.555 & 154.9 & $a=6.5$ kpc, $b=0.26$ kpc\\ 
{\tt Small} & Halo & 0.119 & 72.44 & $r_h=12$ kpc \\ 
\enddata
\end{deluxetable*}

\subsection{Comparison with models}
\label{sec:fitting}

After initializing with the current-day 6D measurements of the Pal~5 system, we model the disruption of the cluster using \textit{galpy}, an orbit-integration-based approximation to calculate action-angle coordinates described in Appendix A of \citet{Bovy_14}. We produce both a leading and a trailing stream, assuming 5 Gyr since the start of disruption (the default in \textit{galpy}). Given the relatively short length of the observed stream ($\sim 22\degree$), which is probably mainly limited by SDSS coverage, we do not attempt to constrain the disruption time. However, in our tests we find that the disruption time has no influence on the specific stream observables used below.

We also do not attempt to fit the radial velocity dispersion of the progenitor, $\sigma_\mathrm{Pal5}(rad)$. $\sigma_\mathrm{Pal5}(rad)$ has the largest influence on the offset 
of the stream relative to the cluster at points closest to the cluster center.
The influence of $\sigma_\mathrm{Pal5}(rad)$ on the stream at further distances from the cluster center is relatively small. 
Therefore, we keep the disruption time and  the dispersion fixed and explore other areas of parameter space: the halo mass, which is connected with
the model parametrization: {\tt Standard}, {\tt Massive} or {\tt Small}, $R_0$, the velocity of the Sun in the Galactic plane ($V_\mathrm{Sun}(\phi)$),  
 the distance of Pal~5, and the proper motion. (The position and radial velocity of the cluster and  the solar parameters 
 $z_\mathrm{Sun}$, $V_\mathrm{Sun}(r)$, and $V_\mathrm{Sun}(z)$ 
 are well-known in comparison, and their errors affect our analysis here negligibly.) 
One finding in our work 
and that of \citet{Pearson_14}, \citet{Kuepper_15} is that when the proper motions in the two dimensions are changed by different multiples of their errors, the resulting stream has a different shape than what is observed.
Therefore, when we change the proper motion, we increase (or decrease) the proper motion in both dimensions by a common multiple of their errors. 

 We find, as expected, that all these parameters make a difference, but three of them are clearly more important than the rest: the halo mass, the distance of Pal~5, and the proper motion.

 \begin{figure}
\begin{center}
\includegraphics[width=0.70 \columnwidth,angle=-90]{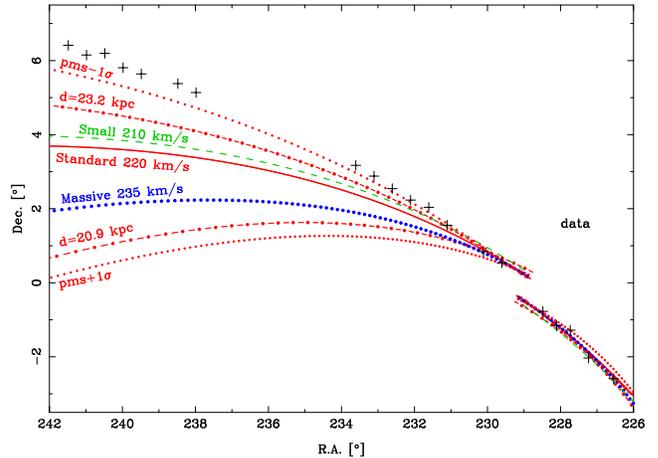} 
\caption{
Comparison of the observed stream streak with different models.
The crosses mark the location ($\alpha, \delta$) of the observed stream debris (Figure~\ref{fig:stream_streak}).
The lines show the different models. When not indicated otherwise they use the standard parameters, $V_0=220$ km/s, 22.05 kpc, and the standard proper motion (pm) of -2.30/-2.26 mas/yr. In the two cases shown, both dimensions of the proper motion have been changed equally in the same direction. The details of the model parameters are explained in the text.
} 
\label{fig:streak_comp}
\end{center}
\end{figure}

In Figure~\ref{fig:streak_comp} we compare the stream streak with the output of the stream disruption models. Specifically, the interpolated track location is what is plotted here as the output of the stream disruption model. 
Due to its shorter length, the leading stream is  much less constraining than the trailing stream.
As discussed, the explored solar parameters are not important for the stream streak: when we decrease $R_0$ and  $V_\mathrm{Sun}(\phi)$
both by 1 $\sigma$, the streak moves up only by 0.4$\degree$ at the very end of the stream  (R.A.$=241\degree$). 
The {\tt Standard} model with $V_0=$220 km/s and $V_\mathrm{20}=$192 km/s produces a trailing stream which clearly has lower Declination values than the observed one.  
A smaller  $V_\mathrm{20}$,  as in the {\tt Small} model, decreases the discrepancy with the data. However, our {\tt Small} model is still clearly offset from the stream.
Even smaller $V_\mathrm{20}$ values are unlikely based on constrains from other data \citep{Sakamoto_03,Koposov_10,Kafle_12,Kuepper_15}. While we cannot exclude that other changes in the details of the potential setup, such as flattening, can result in a good fit, 
we now explore the impact of the phase space parameters of Pal~5 itself: its proper motion and its distance.

We find that larger (more negative) proper motions and/or larger values of distance obtain better consistency with the stream streak, see Figure~\ref{fig:streak_comp}.
Both imply that a larger velocity for Pal 5 in physical units fits the data better.
Good agreement is achieved for either a distance of 24.2 kpc or a change in the proper motion of about $-1.5\,\sigma$.
Because the distance to the cluster has no formal uncertainty, it is difficult to assess the significance of changes in the distance. Currently, both the lower bound and the upper bound of the distance we use are equally likely. Therefore, given that the mean value of our measured proper motion requires larger distances in order to fit the stream better, we interpret this as evidence for a larger Pal~5 distance.
The fact that the {\tt Small} models fit the data better can be interpreted as a preference for a small $V_\mathrm{20}$ and thus a small halo mass.
However, since other parameters, i.e., distance and proper motion, have more leverage within their uncertainties, we do not want to overstate this preference, 
especially in light of adding further stream constraints to the modeling, as discussed below.

 \begin{figure}
\begin{center}
\includegraphics[width=0.70 \columnwidth,angle=-90]{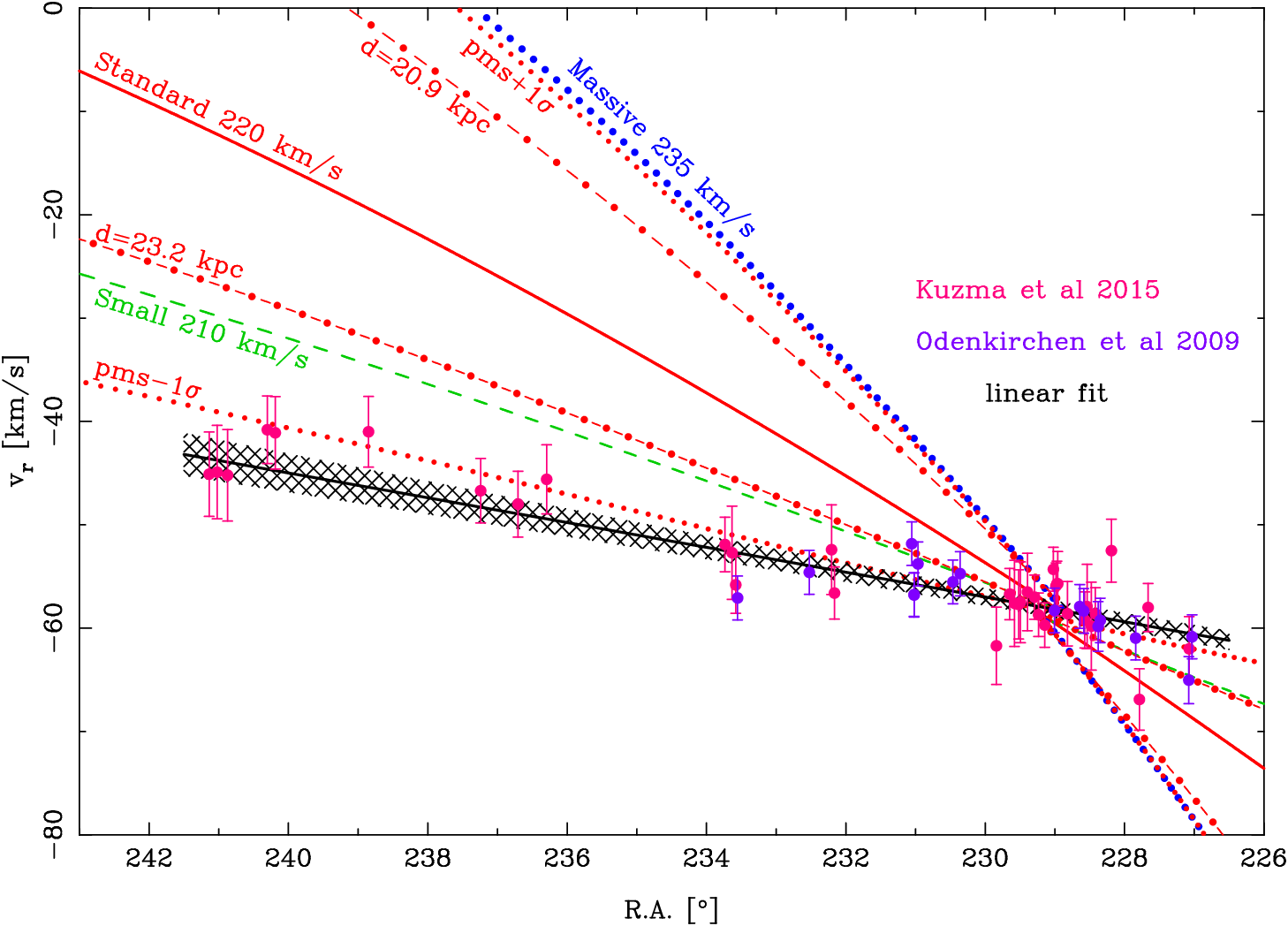}
\caption{We show all velocities of the stream members from \citet{Odenkirchen_09} (violet) and \citep{Kuzma_14} (magenta). For the latter, we do not show the targets that overlap with \citet{Odenkirchen_09}. The errors include a contribution from the measurement errors as well as the dispersion of the stream. We calculate the gradient as function of R.A. from these data points: black line with hatched $1\,\sigma$ error area.
The colored lines show the different models. When not indicated they use the standard parameters of $V_0=220$ km/s, 22.05 kpc and -2.30/-2.26 mas/yr. For the two labeled cases, both dimensions of the proper motion have been changed equally in the same direction. The details are explained in the text.
}
\label{fig:rv_comp}
\end{center}
\end{figure}

The observed radial velocity gradient along the stream provides an additional constraint. In Figure~\ref{fig:rv_comp} we show the corresponding agreement in this space. 
Again, solar parameters are not important, a reduction by 1 $\sigma$ reduces the velocity only by 3 km/s at R.A.$=241\degree$. 
The important parameters are found to be similar to the stream streak case, i.e., cases which fit the stream streak well also fit the radial velocity gradient well. 

Again either a larger distance, or larger (more negative) proper motion are preferred. The best fitting values this time are similar as in the stream streak case, about -1.3$\,\sigma$ 
for the proper motion and 24.2 kpc for the distance. Thus, the {\tt 
Standard} model can fit all the data with a distance of 24.2 kpc. This is on the high end of the currently measured distance, and since the published distance values do not report errors, it is difficult to make a more quantitative statement here.
In the case of changing the proper motion the fit is somewhat worse, 
because a different proper motion is necessary to fit the radial velocities versus the stream streak, while a single distance can fit both equally well.

In Figure~\ref{fig:streak_comp}~and~\ref{fig:rv_comp} it is also visible that the {\tt Small} model can fit both data sets better for the standard distance, although it is a worse fit than changing either proper motion or distance values. 
To quantify this we calculate the $\chi^2$ of the fit to both the stream streak and the radial velocity gradient for several parameter choices. The $\chi^2$  is calculated as follows. For the stream streak we calculate, for each data point, the smallest distance to each model streak and use the distance to calculate the $\chi^2$ for each data point using our error in the stream position. 
Similarly, for the radial velocity,  we calculate for each star the smallest distance to the model gradient to get the $\chi^2$ for each star. The radial velocity error consists of the measurement error of each star and the stream dispersion of 2.1 km/s added in quadrature.
 
\begin{figure}
\begin{center}
\includegraphics[width=0.70 \columnwidth,angle=-90]{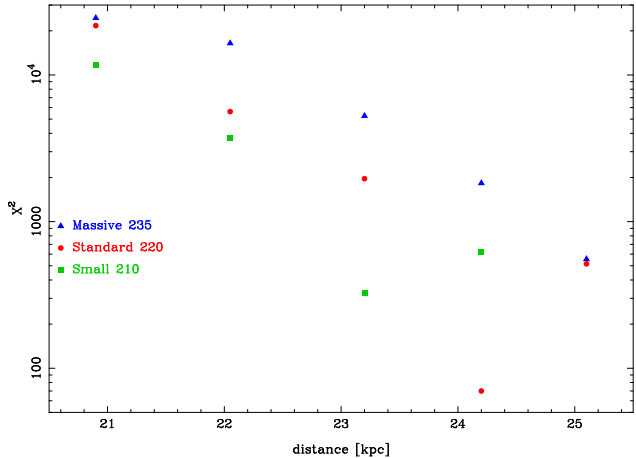} 
\caption{$\chi^2$, shown as a function of distance, for our three models. All models use our measured proper motion of -2.30/-2.26 mas/yr.
The calculation of the $\chi^2$ uses both the distance from the stream streak and from the stream radial velocities.
}
 \label{fig:chi_comp}
\end{center}
\end{figure}

In Figure~\ref{fig:chi_comp} we show the total $\chi^2$ as a function of distance for our three Milky Way models. The smallest $\chi^2$ value (of 70/74) is that 
 of the {\tt Standard} model at 24.2 kpc. The $\chi^2$ values for the other models are higher because in these cases different distances are necessary to minimize the $\chi^2$ for the stream streak versus that of the radial velocity. This  effect causes the {\tt Small} model to fit worse than the {\tt Standard} model. 
When varying the distance, the {\tt Small} model fares better than the {\tt Massive} model.  The reverse is true when we vary the proper motion, even to 
very large values. Thus, both {\tt Small} and {\tt Massive} models are approximately equivalent, and lead to worse fits than the {\tt Standard} model.
As one additional test on the dependence of the details of the Galaxy model we use the interpolated centerline best fit values of \citet{Kuepper_15} (Column 4 of Table 6)\footnote{We choose that model over the other models of this work because in \textit{galpy} only spherical NFW halos are currently implemented.}. 
 This work uses a similar disk and bulge parametrization as used here, but an NFW halo. They marginalize over proper motion and distance, obtaining best-fit values of 23.12 kpc and -2.50/-2.44 mas/yr. When using exactly these parameters (NFW halo, distance and proper motion),  we obtain the lowest $\chi^2$ of all investigated models, of 65.8/72, even when we add to this $\chi^2$ another term that takes into account the difference between the model proper motion and our measured proper motion, which results in a total $\chi^2$ of 68/74. To remind the reader, this is slightly better than the $\chi^2$ of {\tt Standard} model with 24.2 kpc.

 Overall we find that none of the tested models can fit the data, even approximately well, for the standard distance of 22.05 kpc and proper motion. 
Varying both quantities, and within the limitations of our Milky Way model search, we find two Milky Way models that fit the data well: our {\tt Standard} model which has V$_0=220$ km/s and V$_\mathrm{20}=192$ km/s, and one model of \citet{Kuepper_15}, which has 
V$_0=238$ km/s and V$_\mathrm{20}=236$ km/s.
A more thorough Milky Way model search would be easier if a better distance to Pal 5, with an error estimate, were available.

The constraints on some orbit parameters are better: the three best fitting cases, large distance, large absolute proper motion, both with {\tt Standard} model, and interpolated centerline of \citet{Kuepper_15} with their best-fit parameters, all indicate approximately consistent pericenters between 7.74 and 8.28 kpc, respectively. 
In contrast, we obtain looser constraints on the apocenter because it mainly depends on the current distance of Pal~5, which is itself close to apocenter. We obtain between 17.76 and 19.86 kpc for the apocenters of these three cases, respectively.

\section{Discussion and conclusions}
\label{sec:discussion}

We have measured the proper motion of Pal~5, using SDSS DR9 data and LBT/LBC data, separated by $\sim 15$ years, obtaining a proper motion of $\mu_\alpha=-2.296\pm0.186$ mas/yr and $\mu_\alpha=-2.257\pm0.181$ mas/yr. 
We discuss this first in the context of the disruption of Pal~5.
We obtain a pericenter distance of $\sim 8$ kpc.
\citet{Odenkirchen_03} determine the pericenter of the next passage to be about 6 kpc. \citet{Dehnen_04} use a pericenter of 5.5 kpc in their N-body simulations to predict the fate of Pal~5 during the next orbit, obtaining a passage very close to the disk plane.  Independent of the precise mass and shape of the cluster, they obtain that Pal~5 will be disrupted in the next disk passage.
In their recent modeling \citet{Kuepper_15}
obtain a pericenter of about 7.5 kpc. Therefore, the pericenter of the stream orbit is probably larger than in the simulation of \citet{Dehnen_04}. Thus, tidal effects and shocks will probably be weaker in the next passage, which makes it possible that the cluster will not be disrupted in the next passage.

Now we compare with other works on the Galactic potential.
\citet{Odenkirchen_03} predicted the proper motion of Pal~5 using the radial velocity of the cluster and their stream coverage, which includes the central 10$\degree$ of the stream. They use a distance of 23.2 kpc and a logarithmic (spherical) halo potential (without a disk or bulge), and with a constant $V_0$ of 220 km/s.
From orbit fitting to the stream they obtain  $\mu_{\alpha}/\mu_{\delta}=-2.03/-2.045$ mas/yr, just outside of our 1-$\sigma$ errors.
 \citet{Grillmair_06} predict the proper motion by using the radial velocity of the cluster, the stream streak, $R_0=8.5$ kpc, a distance of 23.2 kpc and the $V_0=$220 km/s  potential of \citet{Allen_91}, which consists of a bulge, disk and spherical halo, very similar to our {\tt Standard} potential.
They  obtain a motion  of $\mu_{\alpha}/\mu_{\delta}=-2.27/-2.19$ mas/yr, within our 1$\,\sigma$ interval.
Both works assume that the stream follows the orbit.
 
We also compare our results with those of \citet{Odenkirchen_09}, who compare their analytic results to a prediction of the stream-orbit offset from a full N-body simulation by 
\citet{Dehnen_04}.
They find that this offset becomes important especially when the radial velocity gradient is also used in the fitting. 
In that respect, Pal~5 is different from GD-1, for which stream-fitting (ignoring stream-orbit offsets) seems to do as well as a full N-body model,
However, that is possibly caused by the potential parametrization used by 
\citet{Koposov_10,Bowden_15}, because whether a stream follows an orbit is a very strong function of the global potential model. 
\citet{Odenkirchen_09} obtain a good match to both, the stream streak and the radial velocities, for $V_0\approx220$ km/s. 
They prefer small values of $V_0$, citing 200 km/s or 180 km/s as a better fit to their methods. Based on our tests here, it is possible that the fact that their potential is spherical explains why they prefer a rather small $V_0$.

\citet{Pearson_14} use a combination of a restricted 3-body approach (the streak-line method) to explore parameter space, followed by more detailed N-body simulations, to study the disruption of Pal~5. They leave the proper motion of Pal~5 as a free parameter, and seek to reproduce both the stream contours and the radial velocity gradient, given either a spherical Milky Way halo model (similar to ours in Section~\ref{sec:gal_model}), or the triaxial halo of \citet{Law_10}. 
In the triaxial case they find a best-fit proper motion of $\mu_\alpha/\mu_\delta=-5.0/-3.7$ mas/yr, inconsistent with our measurement. The best fit proper motions for the spherical halo, however, of $\mu_\alpha/\mu_\delta=-2.35/-2.35$ mas/yr, are very consistent with ours. The conclusion of \citet{Pearson_14}, is that a spherical halo gives a better fit to the stream contours and the radial velocity gradient of \citet{Odenkirchen_09} 
than the triaxial halo of \citet{Law_10}, with the triaxial halo causing the stream to ``fan out" much more than observed. Our measured proper motion is consistent with their predicted motion in such a spherical halo for the Milky Way, as is our own modeling of the disruption of the cluster. However, our results do not rule out the presence of a radial gradient in the halo, wherein it changes from more spherical in the inner parts to more triaxial in the outer parts (see also \citealt{Law_10,Vera_13}). 
 
Very recently, \citet{Kuepper_15} used most Pal~5 observables (except for proper motion), including the stream streak, and the radial velocity gradient data of \citet{Odenkirchen_09} to fit the parameters of the Milky Way halo, the Sun and Pal~5. They 
used the streak-line method, obtaining solar parameters which are very similar to the solar parameters used by us. 
The model setup is also similar to ours except that the halo is parametrized with an NFW potential \citep{Navarro_97}, and the flattening of the halo is left as a free parameter. 
They obtain a proper motion of  $\mu_{\alpha}/\mu_{\delta}=-2.39^{+0.15}_{-0.17}/-2.36^{+0.14}_{-0.15}$ mas/yr consistent with our observations. They also obtain a high distance estimate of $23.58^{+0.84}_{-0.72}$ kpc. As a normalization, they obtain $V_0=233^{+12.7}_{-10.0}$ km/s. The halo is consistent with being spherical with $q_z=0.95_{-0.12}^{+0.16}$.

Since our proper motion is slightly smaller than the prediction of \citet{Kuepper_15}, it will be interesting to use this as a prior to obtain new constraints on $V_0$ and the distance. 
In general, our conclusions are similar to those of \citet{Kuepper_15}, like the preference for a relatively larger distance of Pal~5. However, this similarity may be caused by the fact that the bulge and disk in our works are nearly identical, which has a significant effect in the radial range of Pal~5 ($\sim 7 - 20$ kpc). In particular, both of our studies employ a rather massive and extended disk in comparison to recent works, see e.g. \citet{Klypin_02,Rix_13,McMillan_11} and \citet{Bovy_13}. 
Such a disk not only influences the flattening of the potential (see e.g. \citealt{Koposov_10}) but also the enclosed mass within the range probed by Pal 5. 
A further issue of both studies is that there is currently no good distance measurement for Palomar 5, thus adding another nearly free parameter. A good measurement of the distance of Pal 5 with an error would be valuable.
We will explore the effects of a smaller disk and different parameterizations of the halo, the statistical uncertainties in the solar parameters, and the proper motion and distance of Pal~5, together with a full disruption model, to investigate the disruption of Pal~5 in a subsequent paper. 
\newline  

We are grateful to Jo Bovy for his help and advice on using \textit{galpy}, and for useful comments on the text of the paper. We also thank the referee for constructive comments that helped significantly improve the manuscript.
Based on SDSS-III data. Funding for SDSS-III has been provided by the Alfred P. Sloan Foundation, the Participating Institutions, the National Science Foundation, and the U.S. Department of Energy Office of Science. The SDSS-III web site is http://www.sdss3.org/.
SDSS-III is managed by the Astrophysical Research Consortium for the Participating Institutions of the SDSS-III Collaboration including the University of Arizona, the Brazilian Participation Group, Brookhaven National Laboratory, Carnegie Mellon University, University of Florida, the French Participation Group, the German Participation Group, Harvard University, the Instituto de Astrofisica de Canarias, the Michigan State/Notre Dame/JINA Participation Group, Johns Hopkins University, Lawrence Berkeley National Laboratory, Max Planck Institute for Astrophysics, Max Planck Institute for Extraterrestrial Physics, New Mexico State University, New York University, Ohio State University, Pennsylvania State University, University of Portsmouth, Princeton University, the Spanish Participation Group, University of Tokyo, University of Utah, Vanderbilt University, University of Virginia, University of Washington, and Yale University. 

\bibliography{mspap}

\end{document}